\documentclass[fleqn,twoside]{article}
\usepackage[headings]{espcrc2}

\readRCS
$Id: espcrc2.tex,v 1.2 2004/02/24 11:22:11 spepping Exp $
\usepackage{graphicx}
\usepackage{epsfig}
\usepackage[figuresright]{rotating}

\newcommand{\bsmumu}{$B^0_s\rightarrow\mu^+\mu^-$ }

\title{Prospects for Measuring \bsmumu with the CMS Detector}

\author{
Frank-Peter Schilling (for the CMS collaboration)
\thanks{Talk given at BEAUTY 2005, Assisi (Italy), June 2005}
\address[CERN]{CERN/PH, CH-1211 Geneva 23, Switzerland}
}
       
\runtitle{Prospects for Measuring \bsmumu with the CMS Detector}
\runauthor{F.-P. Schilling}

\begin{document}

\begin{abstract}

The flavor-changing neutral current decay \bsmumu is highly
suppressed in the standard model, but its branching fraction of
$3.4*10^{-9}$ could be significantly enhanced through contributions
from new physics.  At the LHC, this rare decay could be observed for
the first time.  In this contribution, the prospects for measuring
\bsmumu with the CMS detector are presented. In particular,
some aspects of the experimental setup, the first and high
level trigger selections, and the offline analysis are discussed.

\vspace{1pc}
\end{abstract}

\maketitle

%%%%%%%%%%%%%%%%%%%%%%%%%%%%%%%%%%%%%%%%%%%%%%%%%%%%%%%%%%%%%%%%%%%%%%%%%%%%%%%

\section{INTRODUCTION}

At the LHC design luminosity of $10^{34} \rm\ cm^{-2}s^{-1}$, around
$10^6$ pairs of b quarks are produced per second.  This makes the LHC
experiments ATLAS, LHC-B and CMS in principle ideal places to study CP
violation, $B_s^0-\bar{B}_s^0$-mixing and rare B decays. In
particular, the experiments have looked into the prospects for measuring
the purely leptonic decay \bsmumu. In the Standard Model (SM), it is
highly suppressed, since this flavor-changing neutral current (FCNC)
process is forbidden at the tree level and can only proceed through
higher order diagrams. In this paper, we present the prospects of
measuring \bsmumu with the CMS detector, with a focus on
several aspects on the experimental setup and the 
online and offline selections.

\subsection{Motivation}

The SM branching fraction for the decay \bsmumu is only
$\mathcal{B}=(3.42\pm0.54)*10^{-9}$
\cite{limitsold,buras}, but it could be significantly enhanced
through contributions from new physics beyond the Standard Model.

For example, in the minimal super-symmetric standard model (MSSM),
$\mathcal{B}(B^0_s\rightarrow\mu^+\mu^-)\propto (\tan \beta)^6$, which
leads to an enhancement of up to three orders of magnitude compared
with the SM, even in the case of minimal flavor violation (MFV).  An
observation would immediately yield an upper bound on the heaviest
mass in the MSSM Higgs sector. Significant enhancements of the
branching fraction are also expected within minimal super-gravity,
non-MFV or R-parity violating super-symmetric models.

\subsection{Limits from Current Experiments}

Since $B_s$ decays are not accessible at the B-factories, the currently
best limits on the decay \bsmumu are obtained at the TEVATRON
experiments CDF and D0 \cite{cdfpub,d0pub}. The most recent result
from D0 \cite{d0prel} is based on a luminosity of $300 \rm\
pb^{-1}$. With 4 candidate events within a mass window of $\pm90 \rm\
MeV$ and $4.3\pm 1.2$ of background, they obtain a limit of
$\mathcal{B}(B^0_s\rightarrow\mu^+\mu^-)<3.0*10^{-7}$.  The best limit
so far comes from CDF \cite{cdfprel}, which use $364 \rm\ pb^{-1}$ and
obtain $\mathcal{B}(B^0_s\rightarrow\mu^+\mu^-)<1.5*10^{-7}$ for a
mass resolution of $\pm 25 \rm\ MeV$.  These limits are still around
two orders of magnitude above the SM expectation, but are expected
to improve with more luminosity.

%%%%%%%%%%%%%%%%%%%%%%%%%%%%%%%%%%%%%%%%%%%%%%%%%%%%%%%%%%%%%%%%%%%%%%%%%%%%%%%

\section{EXPERIMENTAL ASPECTS}

The analysis of the
\bsmumu decay requires the capability to keep the signal
events while reducing the huge QCD background by many orders of
magnitude. The main ingredients to achieve this are a very good
invariant mass resolution in order to reconstruct the di-muon mass,
muon isolation in both the tracker and the calorimeter, as well as a
precise reconstruction of the secondary vertex from the B meson decay.
Since the main detector component used is the tracker,
we will in the following highlight a few relevant features of the
CMS tracker and track reconstruction.

\subsection{The CMS Tracker}

\begin{figure}
\centering
\epsfig{file=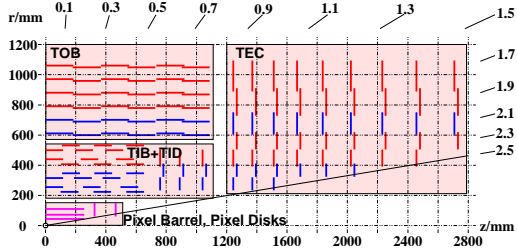,width=0.9\linewidth}
\caption{$rz$ cut through one quarter of the CMS tracker.}
\label{fig:tracker}
\end{figure}

\begin{figure}
\centering
\epsfig{file=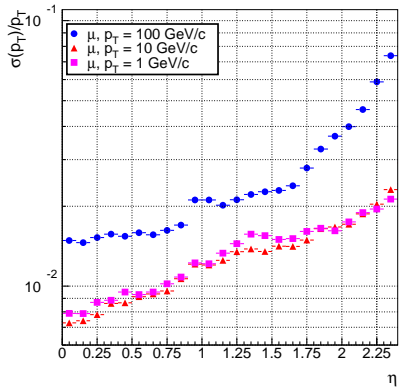,bbllx=72,bblly=166,bburx=539,bbury=625,width=0.7\linewidth}
\caption{$p_T$ resolution of the CMS tracker for single muons.}
\label{fig:ptres}
\end{figure}

The CMS tracker (see Fig. \ref{fig:tracker}) is an all-silicon
detector, which is divided into four strip subdetectors TIB,
TOB (inner and outer barrel), TID (inner discs) and TEC (endcaps), and
two pixel subdetectors (barrel and endcaps), all housed within
a tube of $2.4 \rm\ m$ diameter and $5.4 \rm\ m$ length.

The pixel tracker consists of $66*10^6$ pixels of $100 {\rm \mu m
(r\phi)} \times 150 {\rm \mu m (z)}$ in 1440 modules, arranged
into three barrel layers at $r=4.4,7.3 {\rm \ and \ } 10.2 \rm\ cm$ and two
endcap discs. The hit resolution is $10\ldots 20 \rm\ \mu m$
and the efficiency to find 3 pixel hits per track is $>90\%$ for
$|\eta|<2.2$.  

The strip tracker consists of $>15.000$ strip
modules with a pitch $80\ldots 205 \rm\ \mu m$, distributed over 10
barrel layers (4 TIB, 6 TOB), and 3+9 TID+TEC discs per side.  The
modules of the two innermost layers of TIB and TOB, as well as TID
rings 1,2 and TEC rings 1,2,5 comprise two back-to-back mounted
sensors each, providing a stereo coordinate measurement in both $r\phi$ and
$z$.

The track reconstruction efficiency is $>98\%$ for $|\eta|<2.4$, and
the transverse momentum resolution for single muons with $p_T=1\ldots
100 \rm\ GeV$ is in the range $0.7\ldots 3\%$ for $|\eta|<1.8$ (see
Fig. \ref{fig:ptres}).

\subsection{CMS Alignment Strategy}

In order to fully exploit the potential of the CMS tracker for track
and vertex reconstruction (e.g. for b tagging), the
intrinsic resolution of the silicon strip and pixel sensors of
$10\ldots 20 \rm\ \mu m$ should not be compromised due to imperfect
knowledge of their exact positions and rotations both within the
tracker as well as relative to other parts of the CMS detector, for
example the muon system. Any large misalignments originating from the
initial mechanical construction as well as due to temperature and
dry-out effects during operation must be corrected for. While the
mounting precision of an individual sensor on a module is of
$\mathcal{O}(10\ldots 30 \rm\ \mu m)$, it is only $\mathcal{O}(50\ldots
500 \rm\ \mu m)$ for modules within higher level mechanical structures
(layers, discs).  To correct for these initial misalignments, two
complementary approaches will be implemented: A laser alignment system
(LAS) and several software alignment algorithms using tracks.

The LAS consists of several IR laser beams, which together with
custom alignment position Si sensors are used to continuously monitor
the relative positions of the larger structures of the strip tracker
internally, as well as with respect to the muon system with a
precision of $10\rm\ \mu m$. However, the LAS does not cover the pixel
detector.

The alignment of the pixel tracker, as well as of the individual sensors
of the strip tracker, can only be achieved by running track based alignment
algorithms, using cosmics and beam-halo muons at the LHC
start-up, and muons from $Z^0$ and $W^\pm$ decays during physics data
taking. Due to the large number of more than $16000$ Si sensors, for
which at least six degrees of freedom (3 translations, 3 rotations, 
tilts and sags for composed structures) 
have to be determined, this requires solving a problem
with $\mathcal{O}(100k)$ unknowns. In CMS, several innovative
algorithms are under study at the moment, which try to overcome the
numerically challenging task of inverting very large matrices, for
example using techniques based on the Kalman filter approach.

The CMS alignment strategy currently estimates that at the LHC
start-up the module positions will be known to $\mathcal{O}(100 \rm\
\mu m)$ from the initial placement measurements plus the laser
alignment, which will already make efficient pattern recognition
possible. While the LAS will continuously monitor the composite
structures (except pixel) to $10\rm\ \mu m$, a fast quasi-online 
track based alignment will be implemented to monitor in particular the
pixel detector layers and discs, which is crucial with respect to
the performance of the secondary vertex reconstruction at the High
Level Trigger.  Offline, a full track based alignment at the level of
individual sensors using large samples of tracks will be performed on
a regular basis.

\subsection{Tracking at the High Level Trigger}
\label{sec:hlttracking}

The CMS high level trigger (HLT) consists of a PC farm and reduces the
event rate from $100 \rm\ kHz$ to $100 \rm\ Hz$ by running selection
algorithms on the full event. Due to the limited amount of CPU time
available for the trigger decision, several concepts are foreseen 
to speed up the track reconstruction by reducing the number of
track seeds as well as the number of operations per seed.

Firstly, in {\em regional seed generation} track seeding
is limited to regions of interest (ROI) defined by Level-1 objects,
e.g. a cone around the direction of a muon candidate.

\begin{figure}
\centering
\epsfig{file=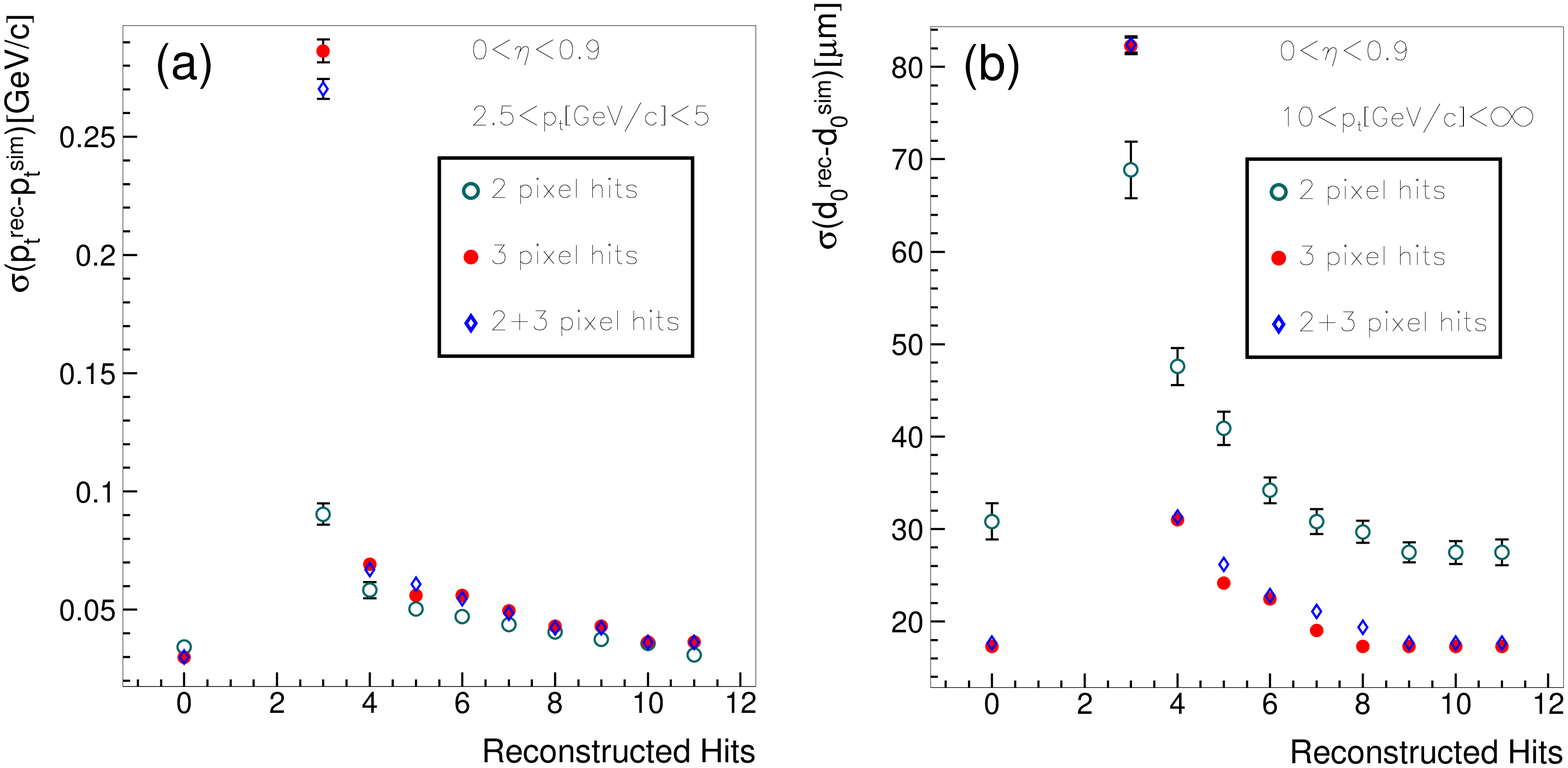,bbllx=560,bblly=165,bburx=1019,bbury=632,width=0.7\linewidth,clip=}
\caption{Impact parameter resolution of the CMS tracker for partial 
and full (shown as ``0 hits'') reconstruction.}
\label{fig:limrec}
\end{figure}

Secondly, in {\em partial} or {\em conditional tracking} 
track reconstruction is stopped if certain criteria are met, such
as the number of reconstructed hits, the transverse momentum
or its resolution have reached given thresholds. It has been
shown \cite{daqtdr} that even when limiting the number of
reconstructed hits to 5 or 6, both efficiency and fake rate are already
similar to the full offline reconstruction performance. The impact
parameter resolution as a function of the number of required hits
is shown in Fig. \ref{fig:limrec}.

%%%%%%%%%%%%%%%%%%%%%%%%%%%%%%%%%%%%%%%%%%%%%%%%%%%%%%%%%%%%%%%%%%%%%%%%%%%%%%%

\section{\bsmumu OFFLINE ANALYSIS}
\label{sec:offline}

The offline analysis of the decay \bsmumu using the CMS detector has
been studied in \cite{cmsnote}. Fully simulated samples of signal and
background Monte Carlo events have been used, the background arising
dominantly from gluon splitting. The
basic kinematic selection cuts applied to the two muon candidates are
transverse momentum $p^\mu_T> 4.3 \rm\ GeV$, rapidity
$|\eta^\mu|<2.4$, distance in $\eta\phi$ between the muons $0.4<\Delta
R_{\mu\mu} < 1.2$ and for the dimuon pair $p^{\mu\mu}_T>12 \rm\
GeV$.  For this kinematic selection, the estimated event numbers for a
luminosity of $10 \rm\ fb^{-1}$ are $N_{signal}=66$ and $N_{bkg}=3*10^{7}$.

A series of further selection cuts are then applied to improve the
signal to background ratio:

\begin{itemize}
\item Dimuon mass window: A cut of $\pm 80 \rm\ MeV$ around the nominal
$B^0_s$ mass of $5.369 \rm\ GeV$ results in a background rejection
of $98.9\%$.

\item Secondary vertex selection: Several cuts are applied on variables
provided by the vertex reconstruction algorithm, such as the minimal
transverse distance between the two muons, the impact parameter and
its error in the transverse plane, and the angle between the secondary
vertex direction and the dimuon momentum in the transverse plane.  The
resulting background rejection power is better than $2.3*10^{-4}$ at
the $90\%$ confidence level. At the same time $~30\%$ of the signal
are kept.

\item Isolation: To further reduce the background, the two muons
are required to be isolated in $r\phi$ in both the tracker as well as
the calorimeter. In the tracker, no charged track with $p_T>0.9 \rm\
GeV$ is required within $\Delta R = 0.5 \Delta R_{\mu\mu} +0.4$.  In
the electromagnetic and hadronic calorimeters the total $E_T$ within
the same cone $\Delta R$ must not exceed 4 (6) GeV in the case of low
(high) lumi. This selection leads to a further background rejection
power of around $0.01$ whilst keeping $45\%$ ($30\%$) of the signal
for low (high) lumi.

\end{itemize}

In summary, these selection cuts reduce the background from $3*10^7$
to $<1$ $(<6.4)$ events for a luminosity of $10 \ (100) \rm\ fb^{-1}$,
whereas the number of signal events amounts to $7 \ (26)$. From these
numbers it follows that a $4\sigma$ observation should be possible
after three years of running at $10 \rm\ fb^{-1}$ per year, even if
the background would be underestimated by a factor two.  However, this
does not yet take into account the first and high level trigger
selections, which are discussed in the next section.

%%%%%%%%%%%%%%%%%%%%%%%%%%%%%%%%%%%%%%%%%%%%%%%%%%%%%%%%%%%%%%%%%%%%%%%%%%%%%%%

\section{\bsmumu ONLINE SELECTION}

\subsection{First Level Trigger}

Purely leptonic B decays in CMS can principally only be triggered using the
single or dimuon trigger at level one, since the electron trigger
thresholds are too high.  The currently foreseen L1 trigger thresholds
for the inclusive isolated muon trigger is 14 GeV, whereas for the
dimuon trigger it is 3 GeV. With these thresholds, the expected
trigger rates are 2.7 and 0.9 kHz, respectively.  Due to the
relatively low threshold of 3 GeV, the dimuon trigger can be used for
the selection of \bsmumu candidates, and it does not cut further into
the basic offline selection as defined in section \ref{sec:offline}.

\subsection{High Level Trigger}

\begin{figure}
\centering
\epsfig{file=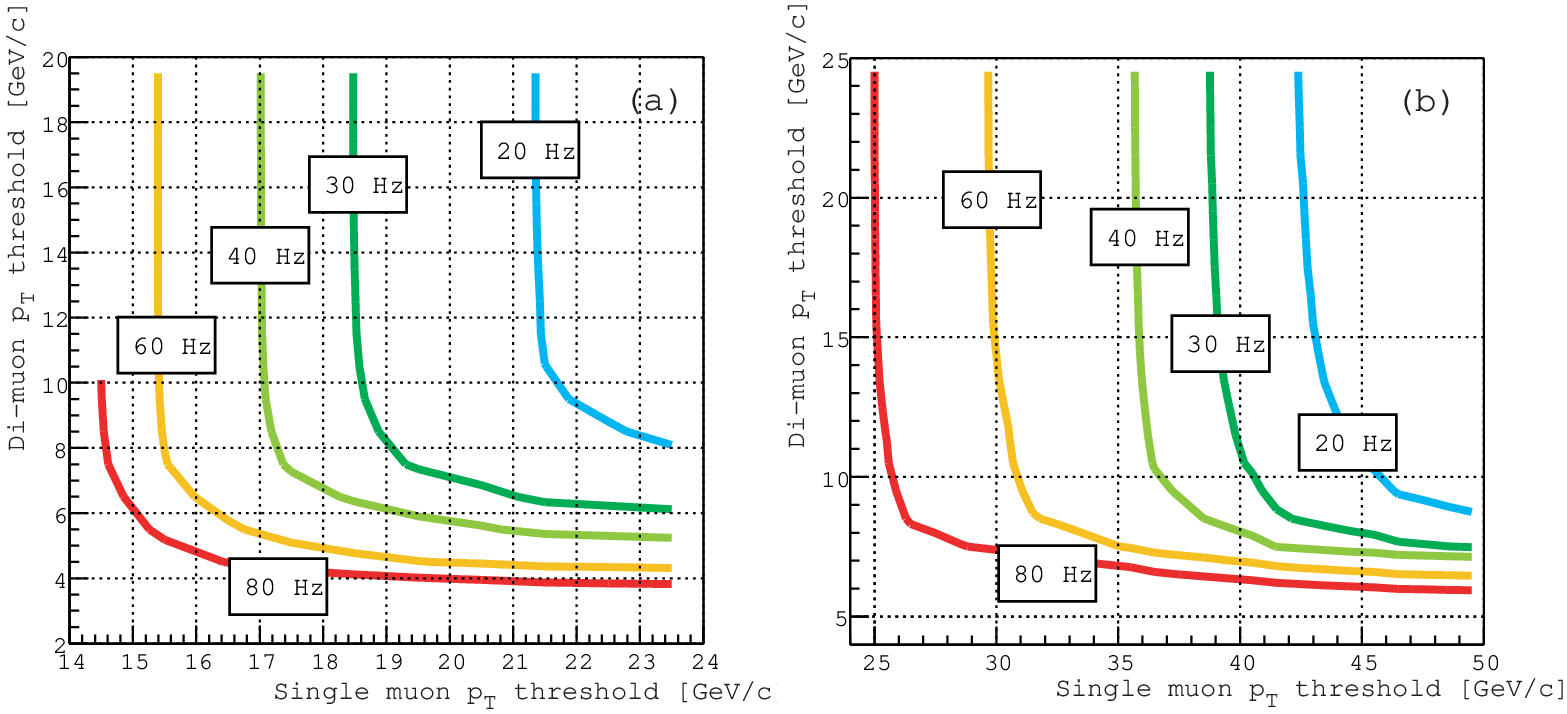,bbllx=0,bblly=275,bburx=315,bbury=555,angle=0,width=0.7\linewidth,clip=}
\caption{Single and dimuon high level trigger rates for low luminosity.}
\label{fig:hltmuons}
\end{figure}

\begin{figure}
\centering
\epsfig{file=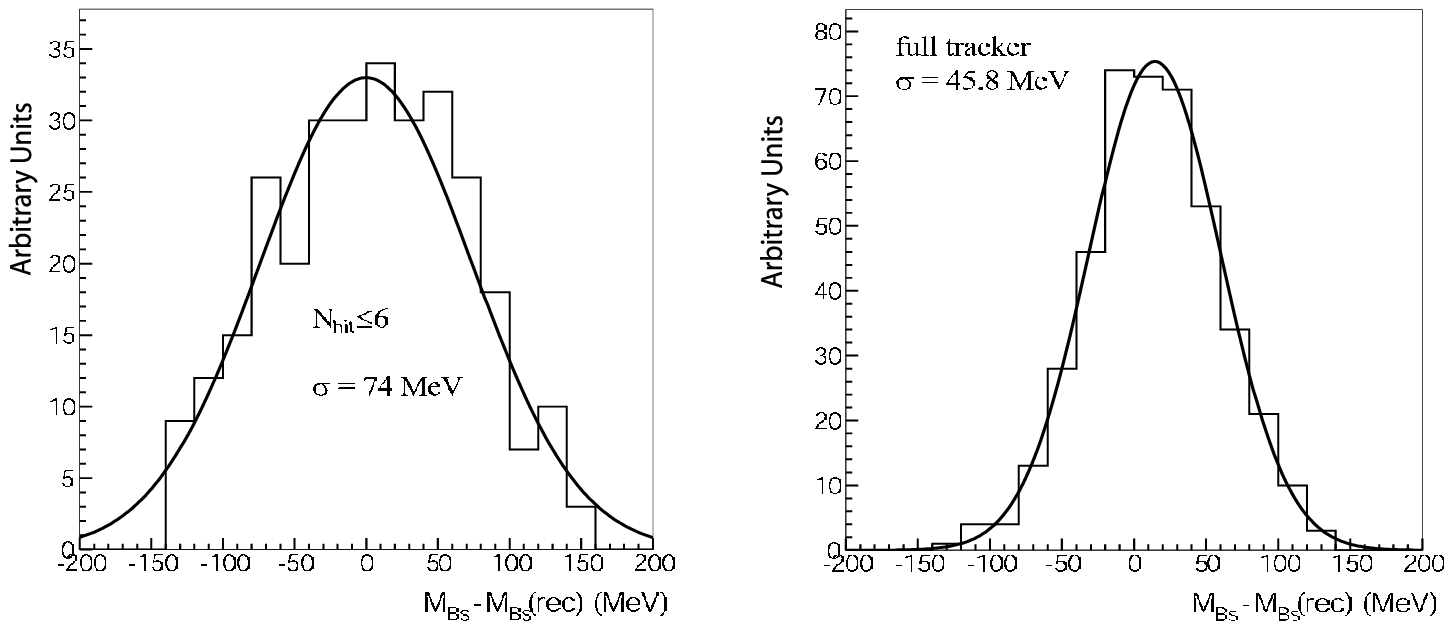,angle=270,width=0.65\linewidth,clip=}
\caption{Dimuon invariant mass of the \bsmumu candidate for HLT 
 reconstruction.}
\label{fig:masshlt}
\end{figure}

\begin{figure}
\centering
\epsfig{file=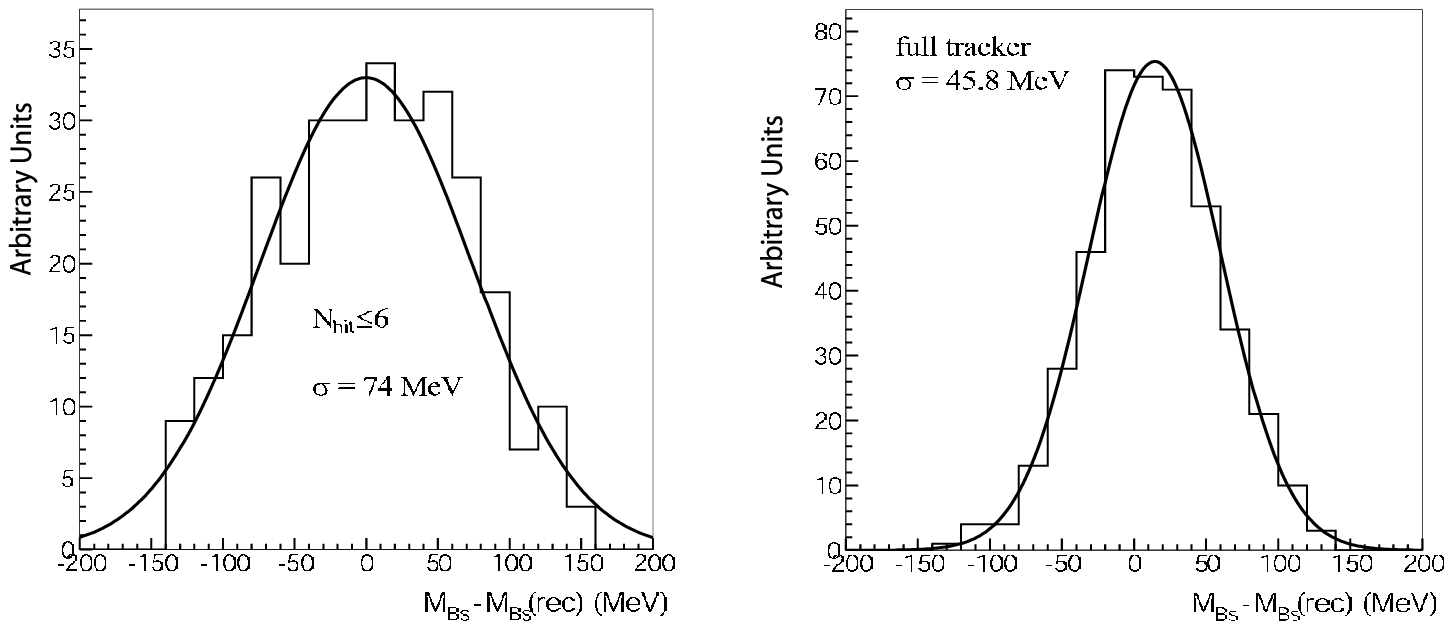,angle=270,width=0.65\linewidth,clip=}
\caption{Dimuon invariant mass of the \bsmumu candidate for 
full reconstruction.}
\label{fig:mass}
\end{figure}

It is currently foreseen that out of the 100 Hz HLT output rate,
approximately 30 Hz are allocated to the inclusive single and dimuon
trigger channels, resulting in trigger thresholds of 19 (7) GeV for
the single (di) muon trigger, respectively (see
Fig. \ref{fig:hltmuons}) \cite{daqtdr}.  The contribution of processes
involving charm and beauty quarks to the single muon rate is only
$\sim 25\%$.  Hence, it is very inefficent to use inclusive muon
triggers for rare B decays, and efficient HLT selection algorithms are
mandatory.

The HLT selection algorithm for the \bsmumu channel \cite{daqtdr}
makes use of both regional and conditional tracking (see section
\ref{sec:hlttracking}). Pixel seeds are only considered
within cones around the L1 muon candidates with $p_T>4 \rm\ GeV$ and
impact parameter $<1 \rm\ mm$, compatible with the primary
vertex. Track reconstruction is stopped if the transverse momentum is
$<4 \rm\ GeV$ with $5\sigma$ significance.  Only tracks with at least
six hits and $\sigma(p_T)/p_T<2\%$ are kept.  If exactly two opposite
sign tracks are found, their invariant mass is calculated and only
pairs with a mass within $\pm 150\rm\ MeV$ around the $B_s^0$ mass
are retained.  They must be compatible with a secondary vertex with an
impact parameter $d_0>150 \rm\ \mu m$ and $\chi^2<20$.  The dimuon
invariant mass resolution is found to be 74 MeV at the HLT
(Fig. \ref{fig:masshlt}), compared with 46 MeV using the full offline
reconstruction (Fig. \ref{fig:mass}).

\subsection{\bsmumu Efficiency and Rates}

For the trigger selection described above, the signal efficiencies are
$15.2\%$ and $33.5\%$ at the first and high level trigger,
respectively, which amounts to a global efficiency of $5.1\%$
\cite{daqtdr}.  This corresponds to $47$ \bsmumu events for a
luminosity of $10 \rm\ fb^{-1}$. The trigger rate (dominated by
background) is below $2 \rm\ Hz$.

%%%%%%%%%%%%%%%%%%%%%%%%%%%%%%%%%%%%%%%%%%%%%%%%%%%%%%%%%%%%%%%%%%%%%%%%%%%%%%%

\section{CONCLUSIONS}

CMS is well suited for b physics in general and
rare B decays in particular, due to the high luminosity,
the precise all-silicon tracker and the
powerful muon system, which also provides a first level trigger.
Crucial ingredients for the selection of the rare decay \bsmumu are a
low transverse momentum muon threshold at L1 and a very efficient
online reconstruction at the high level trigger. Both
secondary vertex and invariant mass reconstruction require that the
very good intrinsic resolution of the CMS silicon tracker is not
significantly compromised due to misalignment.

An observation of the decay \bsmumu can place severe constraints on
extensions of the standard model, such as super-symmetry.  It has been
shown that an observation in the CMS detector within the first few
years of data taking is possible, even with moderate
luminosity. Hence, the \bsmumu channel represents a very promising
topic on the early physics agenda of CMS.

%%%%%%%%%%%%%%%%%%%%%%%%%%%%%%%%%%%%%%%%%%%%%%%%%%%%%%%%%%%%%%%%%%%%%%%%%%%%%%%

%%%%%%%%%%%%%%%%%%%%%%%%%%%%%%%%%%%%%%%%%%%%%%%%%%%%%%%%%%%%%%%%%%%%%%%%%%%%%%%

\end{document}